\documentclass[twocolumn]{aastex7}

\def\lapp{\ifmmode\stackrel{<}{_{\sim}}\else$\stackrel{<}{_{\sim}}$\fi}
\def\gapp{\ifmmode\stackrel{>}{_{\sim}}\else$\stackrel{>}{_{\sim}}$\fi}
\usepackage{graphicx}
\usepackage{multirow}
\usepackage{subcaption}
\usepackage{color}
\usepackage{amsmath}
\usepackage{soul}
\usepackage{hyperref}
\usepackage{lineno}
\newcommand{\fluxcgs}{\ensuremath{\mathrm{erg}\, \mathrm{s}^{-1}\, \mathrm{cm}^{-2}}}
\newcommand{\lumcgs}{\ensuremath{\mathrm{erg}\, \mathrm{s}^{-1}}}
\newcommand{\kms}{\ensuremath{\mathrm{km}\, \mathrm{s}^{-1}}}
\newcommand{\evcm}{\ensuremath{\mathrm{eV}\, \mathrm{cm}^{-3}}}
\newcommand{\fluxtt}{F_{2\text{--}10\,\rm{keV}}}
\def\lapp{\ifmmode\stackrel{<}{_{\sim}}\else$\stackrel{<}{_{\sim}}$\fi}
\def\gapp{\ifmmode\stackrel{>}{_{\sim}}\else$\stackrel{>}{_{\sim}}$\fi}
\newcommand{\marka}{$^{\rm a}$}
\newcommand{\markb}{$^{\rm b}$}
\newcommand{\markc}{$^{\rm c}$}
\newcommand{\markd}{$^{\rm d}$}

\shorttitle{}
\shortauthors{}

\begin{document}

\title{X-ray studies of PSR~J1838$-$0655 and its wind nebula associated with HESS~J1837$-$069 and 1LHAASO~J1837$-$0654u}

\correspondingauthor{Hongjun An}
\email{hjan@cbnu.ac.kr}
\author[0009-0000-4478-6437]{Minseo Park}
\email{}
\affiliation{Department of Astronomy and Space Science, Chungbuk National University, Cheongju, 28644, Republic of Korea}
\author[0000-0002-9103-506X]{Jaegeun Park}
\email{}
\affiliation{Department of Astronomy and Space Science, Chungbuk National University, Cheongju, 28644, Republic of Korea}
\author[0000-0003-0226-9524]{Chanho Kim}
\email{}
\affiliation{Department of Astronomy and Space Science, Chungbuk National University, Cheongju, 28644, Republic of Korea}
\author[0000-0002-6389-9012]{Hongjun An}
\email{hjan@cbnu.ac.kr}
\affiliation{Department of Astronomy and Space Science, Chungbuk National University, Cheongju, 28644, Republic of Korea}

\begin{abstract}
We analyzed X-ray data from Chandra, XMM-Newton, NICER, and NuSTAR to characterize the properties of the pulsar PSR~J1838$-$0655 and its pulsar wind nebula (PWN) associated with HESS~J1837$-$069. Based on 5.5 years of NICER monitoring, we detected a glitch around MJD~59300, characterized by a fractional frequency jump of approximately $2\times 10^{-6}$.
We constructed semi-phase-coherent timing solutions for pre- and post-glitch epochs, allowing for phase alignment of multi-instrument data and a subsequent measurement of the pulsed spectrum of the pulsar. This analysis confirmed previously-reported spectral curvature and revealed a peak energy of $73^{+85}_{-26}$\,keV in the pulsar's spectral energy distribution (SED), based on a {\tt logpar} model fit of the pulsed spectrum. We discuss these findings within the framework of pulsar magnetospheric emission scenarios. The PWN's X-ray spectrum is well-described by a power law with a photon index of $2.1\pm0.3$, softer than previously-reported measurements.
We also characterized the X-ray emission from another extended X-ray source AX~J1837.3$-$0652 within the extent of HESS~J1837$-$069. Based on the spatial and spectral properties of these X-ray sources, we propose a leptonic emission scenario for HESS~J1837$-$069 and demonstrate its feasibility through SED modeling. Finally, we discuss the implications of our model results and alternative scenarios for the gamma-ray emission.
\end{abstract}


\section{Introduction}
\label{sec:sec1}
Rotation-powered pulsars (RPPs) possessing spin-down luminosities of $\dot E_{\rm SD}\gapp 10^{36}$\,\lumcgs\ often exhibit non-thermal X-ray emission and drive extended pulsar wind nebulae (PWNe). These systems provide critical windows into the physical processes governing pulsar magnetospheres \citep[e.g.,][]{Torres2018} and contribute significantly to our understanding of the Galactic population of energetic cosmic-ray leptons \citep[e.g.,][]{Slane2017,Sudoh2023}. 

A subset of these energetic RPPs is characterized by hard X-ray spectra, typically with photon indices $\Gamma<1.5$, broad, single-peaked X-ray pulse profiles, and relatively weak radio and GeV emission. These pulsars, including PSR~B1509$-$58, PSR~J1838$-$0655, PSR~J1846$-$0258, and PSR~J1849$-$0001, have been identified as a distinct population by \citet{Harding2017}.
These authors introduced the term ``MeV pulsars'' to describe these objects, suggesting that their observed properties offer valuable constraints on pair production mechanisms within pulsar magnetospheres.

\citet{Harding2017} proposed that the peak energy ($E_{\rm SR}$) of synchrotron emission in MeV pulsars is dependent on the location of electron-positron pair production and subsequent radiation. They specifically considered polar cap and outer gap scenarios for pair production, with emission occurring in the outer magnetosphere in both cases.
\citet{Kim2024} tested theses scenarios with $E_{\rm SR}$ measurements derived from a log-parabolic model fit for three MeV pulsars \citep[PSR~B1509$-$58, PSR~J1846$-$0258, and PSR~J1849$-$0001;][]{Chen2016, Kuiper2018,Kim2024}. Their analysis suggested that the outer gap scenario is disfavored because the observed trend between $E_{\rm SR}$ and the magnetic field at the light cylinder ($B_{\rm LC}$) was opposite to the prediction of this scenario. In contrast, the polar cap scenario remained viable. Given the simplified nature of these models, a larger pulsar sample is essential to explore more sophisticated physical scenarios.

The PWNe associated with MeV pulsars are also intriguing as they are suggested to be sites of extreme particle acceleration, reaching PeV energies. SED modeling of Kes~75, associated with PSR~J1846$-$0258, suggests leptonic PeV acceleration \citep{Gelfand2014}.  LHAASO detection of ultra high-energy (UHE; $>100$\,TeV) emission from the PWN of PSR~J1849$-$0001 \citep[][]{Cao2024} provides direct evidence for UHE acceleration, with an SED study indicating electron energies of $\sim 4$\,PeV \citep[e.g.,][]{Zhu2024}, comparable to or higher than other leptonic PeVatron candidates \citep[e.g.,][]{Woo2023}.

PSR~J1838$-$0655 (J1838 hereafter) is a pulsar characterized by a 70.5-ms spin period ($P$), its derivative $\dot P=4.9\times 10^{-14}$, and a spin-down power of $\dot E_{\rm SD} = 5.5 \times 10^{36}$\,\lumcgs\ \citep{Gotthelf2008}. It exhibits properties typical of MeV pulsars, including a hard X-ray spectrum ($\Gamma<1.5$) and a broad pulse profile, leading to its classification as such by \citet{Harding2017}.  However, the X-ray SED of J1838 remains a subject of ongoing investigation.
\citet{Lin2009} analyzed Chandra, RXTE, and Suzaku observations, favoring a broken power law (BPL) over a simple power law (PL). In contrast, \citet{kh15} fit RXTE and INTEGRAL data with a log-parabolic model.  More recently, \citet{Takata2024} confirmed and refined the BPL results of \citet{Lin2009} using NICER, XMM-Newton, and NuSTAR observations. While the BPL model adequately represents the X-ray spectra, the possibility of a high-energy cutoff, which allows for an estimation of $E_{\rm SR}$, warrants further consideration.
The log-parabolic model can be refined utilizing the enhanced sensitivity of NICER, XMM-Newton, and NuSTAR over a broad X-ray band, providing an opportunity for a precise estimation of $E_{\rm SR}$.

\begin{figure}[t]
\centering
\includegraphics[width=3.3 in]{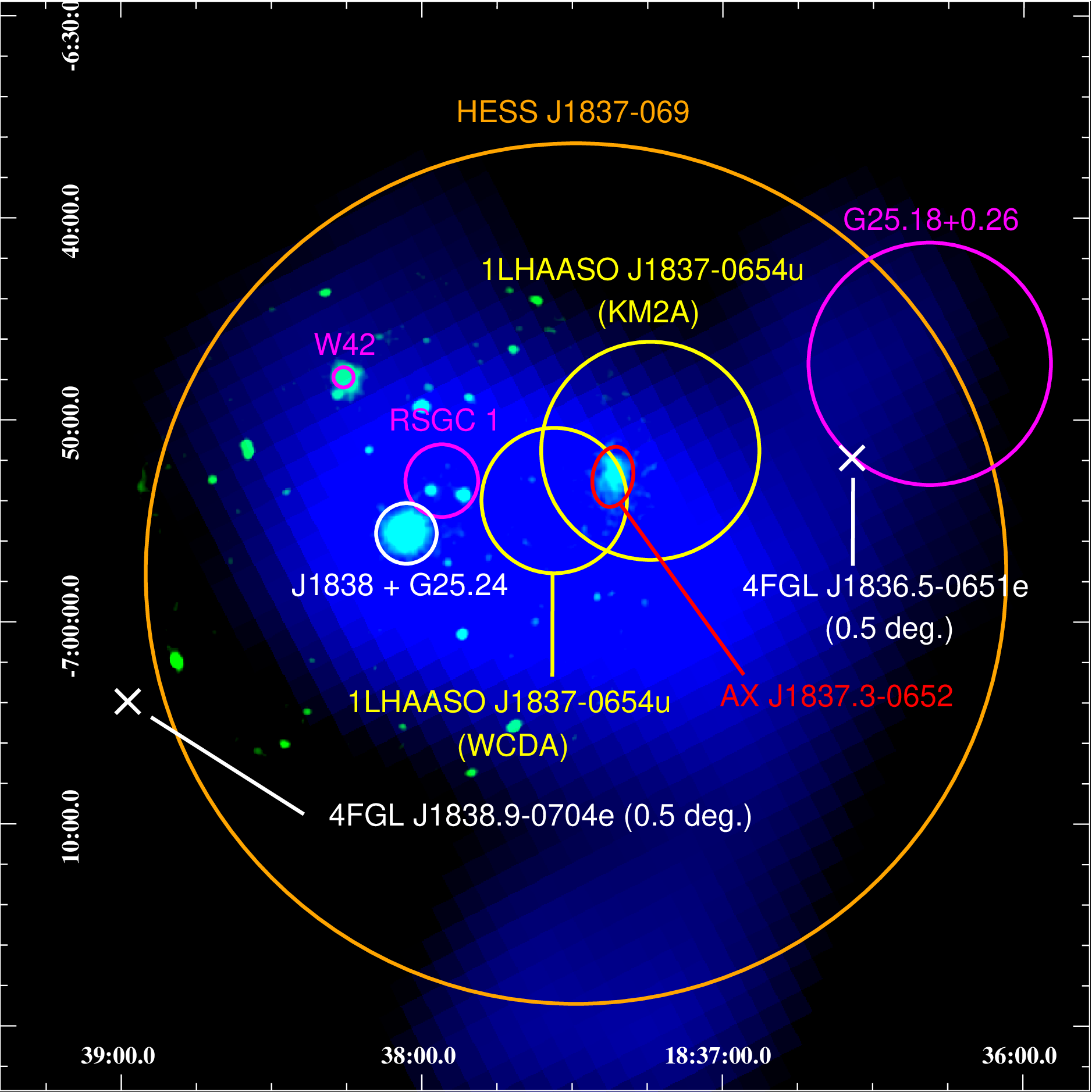} \\
\caption{Composite image of the HESS~J1837$-$069 field constructed using XMM-Newton MOS (green) and H.E.S.S. (blue) data \citep[][]{HESSHGPS2018}. The white circle ($R = 90''$) indicates the J1838+G25.24 system, and the red ellipse ($60'' \times 90''$) marks the candidate PWN AX~J1837.3$-$0652 (AX~J1837). Gamma-ray emission regions are also shown; the white crosses (Fermi-LAT) mark the centers of extended emissions with $1\sigma$ widths of $0.5^\circ$ \citep[][]{Ballet2023}, the orange circle ($R=0.355^\circ$) shows the $1\sigma$ width of HESS~J1837$-$069, and the yellow circles display the positional uncertainties of the point source 1LHAASO J1837$-$0654u \citep[][]{Cao2024}. Magenta circles denote the star clusters W42, RSGC~1, and G25.18+0.26 \citep[][]{Katsuta2017,Sun2020}.}
\label{fig:fig1}
\vspace{0mm}
\end{figure}

G25.24$-$0.19\footnote{http://snrcat.physics.umanitoba.ca/SNRtable.php} (G25.24 hereafter), the PWN surrounding J1838, is also an intriguing object due to its potential association with HESS~J1837$-$069 \citep[e.g.,][]{Gotthelf2008} and 1LHAASO~J1837$-$0654u \citep[][]{Cao2024}. However, definitive association remains challenging given the large extent of HESS~J1837$-$069 and potential very high-energy (VHE; $>100$\,GeV) and UHE emissions from other sources within the region (Figure~\ref{fig:fig1}), including star clusters and an extended X-ray source \citep[e.g.,][]{Kargaltsev2012}. The region's complexity is further evidenced by the Fermi large area telescope \citep[LAT;][]{Atwood2009} and H.E.S.S. observations, which necessitate two- and three-component models, respectively \citep[][]{Ballet2023, HESSHGPS2018}. Consequently, previous studies \citep[e.g.,][]{Katsuta2017,Sun2020} proposed multi-component models, incorporating G25.24 and young star clusters to explain the GeV-TeV emission. However, the absence of X-ray and UHE data in these analyses necessitates a refined SED study to better constrain the region's emission mechanisms.

Given that the highest-energy electrons, responsible for VHE and UHE emissions, also produce X-rays, detailed X-ray characterization is crucially important for understanding the origin of the VHE and UHE emissions. However, the current spectral determination for G25.24 \citep[][]{Gotthelf2008} suffers from significant uncertainty, precluding a detailed SED study for HESS~J1837$-$069. Furthermore, the PWN candidate AX~J1837.3$-$0652 (hereafter AX~J1837; Figure~\ref{fig:fig1}), located $\sim10'$ west of G25.24 and within HESS~J1837$-$069, may also contribute to the observed VHE \citep[e.g.,][]{Gotthelf2008,Kargaltsev2012} and UHE emissions. These factors hinder a definitive identification of the origin of these emissions.

In this paper, we present a comprehensive analysis of J1838, its PWN G25.24, and AX~J1837 using X-ray observations by NICER, XMM-Newton, Chandra, and NuSTAR (Sections \ref{sec:sec2}).  We perform a timing analysis to establish a long-term timing solution for the pulsar, and we characterize the morphology of G25.24. These results are used to investigate the spectral properties of the pulsar, G25.24, and AX~J1837. We then employ a multi-zone PWN model to explore whether these two extended X-ray sources can account for the broadband emission observed from the HESS~J1837$-$069 region (Section \ref{sec:sec3}). Finally, we interpret our findings in the context of MeV pulsars and X-ray PWNe, and discuss alternative scenarios for HESS~J1837$-$069 (Section \ref{sec:sec4}). Uncertainties reported in this paper are $1\sigma$ confidence intervals unless otherwise noted.

\begin{table}[t]
\begin{center}
\caption{Data used in this work}
\label{ta:ta1}
\scriptsize{
\hspace{-15 mm}
\begin{tabular}{lcccc} \hline\hline
Instrument & Obs. ID(s) & Date (MJD) \\ \hline
Chandra  & {6719, 16673}\marka & 53967, 57432 \\
XMM  & 0552950101, 0852240101 & 54756, 58775 \\
NICER & 1020680101--4607011906 & 58268--59764 \\
NuSTAR  & 30501013002, 30501013004 & 58775, 58778 \\\hline
\end{tabular}}
\end{center}
\vspace{-2.0 mm}
\footnotesize{
\marka{Data contained in \href{https://doi.org/10.25574/cdc.357}{Chandra Data Collection (CDC) 357}.}}
\end{table}

\section{X-ray Data Analysis}
\label{sec:sec2}

\subsection{Data Reduction}
\label{sec:sec2_1}

\begin{figure}
\centering
\hspace{-9.5mm}
\includegraphics[height=2.39 in]{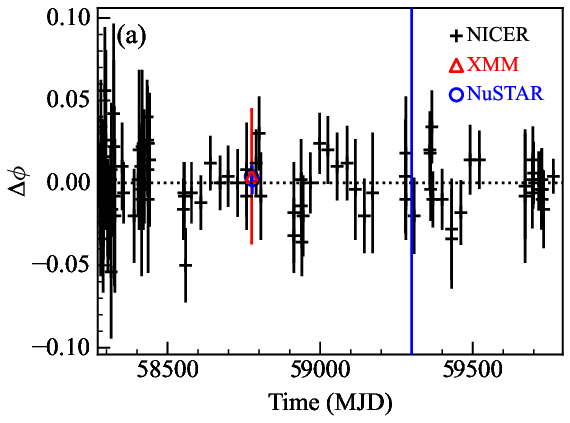} \\
\hspace{-4mm}
\includegraphics[height=2.4 in]{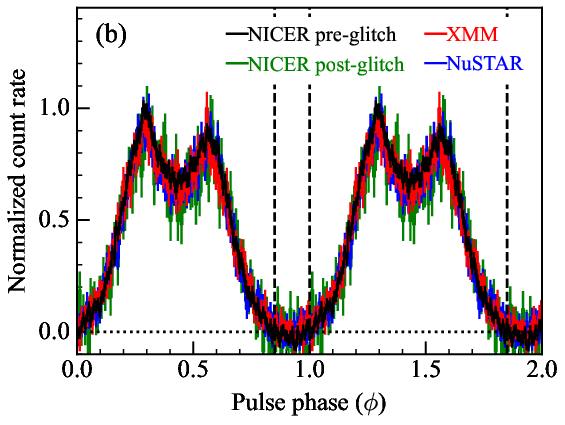} \\
\caption{Timing analysis results for J1838.
(a) Phase residuals of the NICER (black), XMM-Newton PN (red), and NuSTAR (blue) data after applying the timing solutions presented in Table~\ref{ta:ta2}. The blue vertical line (MJD~59300) marks the glitch epoch. (b) Pulse profiles measured by NICER (pre-glitch: black, post-glitch: dark green), XMM-Newton PN (red), and NuSTAR (blue), generated using the timing solutions from Table~\ref{ta:ta2}. Off-pulse emission has been subtracted and the profiles are normalized to 1 at their maximum. The vertical dashed lines mark the off-pulse phase interval ($\phi=0.85\textrm{--}1$).
\label{fig:fig2}}
\vspace{0mm}
\end{figure}

We analyzed X-ray data of the HESS~J1837$-$069 region obtained by Chandra, XMM-Newton, NICER, and NuSTAR (Table~\ref{ta:ta1}). Chandra ACIS-I observations (Obs. IDs 6719 and 16673 taken on 2006 August 19 and 2016 February 13, respectively) were reprocessed using the \texttt{chandra\_repro} tool (CIAO 4.15), yielding net exposure times of 20\,ks and 54\,ks, respectively.  XMM-Newton observations (Obs. IDs 0552950101 and 0852240101) were conducted on 2008 October 6 and 2019 October 19. The MOS detectors were operated in full-frame mode, while the PN detector used small-window mode. Data reduction was performed using the \texttt{emproc} and \texttt{epproc} tools (SAS 20230412\_1735), and particle flares were removed following standard procedures.  For Obs. ID 0852240101, the resulting net exposure times were 61\,ks (MOS1), 66\,ks (MOS2), and 62\,ks (PN).  For Obs. ID 0552950101, the net exposure times were 59\,ks (MOS1), 59\,ks (MOS2), and 44\,ks (PN). 

We used HEASOFT v6.33 to analyze the NICER and NuSTAR data. 113 NICER observations spanning from 2018 May 29 (Obs. ID 1020680101) to 2022 July 4 (Obs. ID 4607011906) were reprocessed using the \texttt{nicerl2} script. Some NICER observations had very short exposure times, precluding meaningful timing or spectral analysis. NuSTAR observations (Obs. IDs 30501013002 and 30501013004 taken on 2019 October 17 and 21, respectively) were processed using the \texttt{nupipeline} tool with optimized SAA filtering.  The resulting net exposure times were 113\,ks for the 2019 October 17 observation and 53\,ks for the 2019 October 21 observation.

\subsection{Pulsar Timing Analysis}
\label{sec:sec2_2}
To derive a long-term timing solution for phase-resolved spectroscopy, we conducted a timing analysis utilizing high timing resolution data from XMM-Newton PN, NICER, and NuSTAR. Source events were extracted in the 2--10\,keV (XMM-Newton PN), 2--8\,keV (NICER), and 3--60\,keV (NuSTAR) bands. The XMM-Newton and NuSTAR data were further filtered spatially using circular regions of $R=16''$ and $R=60''$, respectively. We then applied a barycentric correction to the event arrival times using the source position (R.A., decl.)=($279.51304^\circ$, $-6.92594^\circ$).

\begin{table}[t]
\vspace{-0.0in}
\begin{center}
\caption{Timing parameters of PSR~J1838-0655}
\label{ta:ta2}
\vspace{-0.05in}
\scriptsize{
\hspace{-8 mm}
\begin{tabular}{lc} \hline\hline
Parameter				& Value   \\ \hline
 \multicolumn{2}{c}{Pre-glitch timing solution}\\
Range of dates (MJD)			& 58268--59281   \\
Epoch (MJD TBD)				& 58268.99999979   \\
Frequency (Hz)				& 14.181573880(4) \\
1st derivative (Hz $\rm s^{-1}$)	& $-9.978(1)\times 10^{-12}$  \\
2nd derivative (Hz $\rm s^{-2}$)	& $1.2(1)\times 10^{-21}$   \\
3rd derivative (Hz $\rm s^{-3}$)	& $-7(1)\times 10^{-29}$   \\
4th derivative (Hz $\rm s^{-4}$)	& $3.5(6)\times 10^{-36}$  \\
5th derivative (Hz $\rm s^{-5}$)	& $-8(1)\times 10^{-44}$   \\ \hline
 \multicolumn{2}{c}{Post-glitch timing solution}\\
Range of dates (MJD)			& 59308--59764\\
Epoch (MJD TBD)				& 59358.99999963 \\
Frequency (Hz)				& 14.180665582(3) \\
1st derivative (Hz $\rm s^{-1}$)	& $-1.0015(1)\times 10^{-11}$\\
2nd derivative (Hz $\rm s^{-2}$)	& $1.8(2)\times 10^{-21}$\\
3rd derivative (Hz $\rm s^{-3}$)	& $-7(1)\times 10^{-29}$\\
\hline
\end{tabular}}
\end{center}
\footnotesize{Note. Values in parentheses represent $1\sigma$ uncertainties.}
\vspace{-2.0 mm}
\end{table}

Our initial analysis of individual XMM-Newton, NICER, and NuSTAR observations employed the H-test \citep{drs89} to search for pulsations around a spin frequency $\nu$ of 14.181\,Hz, while holding the frequency derivative fixed at $-9.909\times10^{-12}\ {\rm Hz\,s^{-1}}$ \citep{Gotthelf2008}. Despite significant scatter in the derived $\nu$ values, largely attributed to flares and data gaps in the NICER data, we successfully detected pulsations within the $\nu\sim 14.180$\,Hz--$14.182$\,Hz range, which showed a gradual decrease with time. This analysis further revealed a frequency jump (i.e., glitch) of a fractional size $\Delta\nu/\nu \sim 2 \times10^{-6}$ and a possible change in $\dot \nu$ around MJD 59300 (see below).

\begin{figure}
\centering
\includegraphics[height=2.4 in]{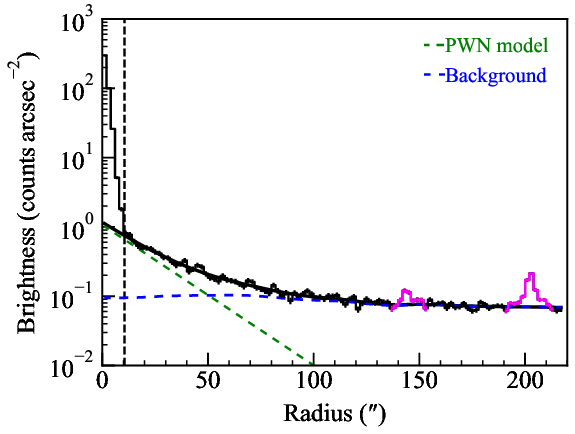} 
\caption{1--8\,keV radial brightness profile of the J1838+G25.24 system (black histogram) and model fit (solid black curve). The fit excludes the central $10.5''$ (vertical dashed line) to avoid pulsar contamination and regions with bright sources (magenta points). The dashed green curve represents the PWN model for G25.24, while the dashed blue curve shows the background emission. \label{fig:fig3}}
\vspace{0mm}
\end{figure}

While this initial approach provided a useful overview, we employed a semi-phase-coherent analysis \citep[e.g.,][]{AnArchibald2019} to derive accurate timing solutions and phase-align the pulse profiles from the multi-instrument data. To mitigate the effects of timing noise, higher-order time derivatives of the spin frequency were incorporated. The high cadence of the NICER data enabled phase connection, facilitating precise phase alignment of the pulse profiles. However, due to the glitch, phase connection could not be maintained across the glitch epoch. Consequently, we constructed separate timing solutions for the pre- and post-glitch periods. The resulting phase residuals and pulse profiles are displayed in Figure~\ref{fig:fig2}a and b. The pre- and post-glitch solutions, presented in Table~\ref{ta:ta2}, robustly verify the glitch and the change of $\dot \nu$ found in our initial analysis.

Based on the pulse profile, we defined the on-pulse and off-pulse phases as $\phi=0\textrm{--}0.85$ and $0.85\textrm{--}1$, respectively. The on-pulse profiles (with off-pulse emission subtracted; Figure~\ref{fig:fig2}b) measured by the three observatories show excellent agreement. The 2008 XMM-Newton observation could not be phase-connected and was excluded from this timing analysis. However, its phase was shifted to align its profile with those in Figure~\ref{fig:fig2}b for spectral analysis in Section~\ref{sec:sec2_4_1}.

\begin{figure*}[t]
\centering
\begin{tabular}{ccc}
\hspace{-2 mm}
\includegraphics[width=2.325 in]{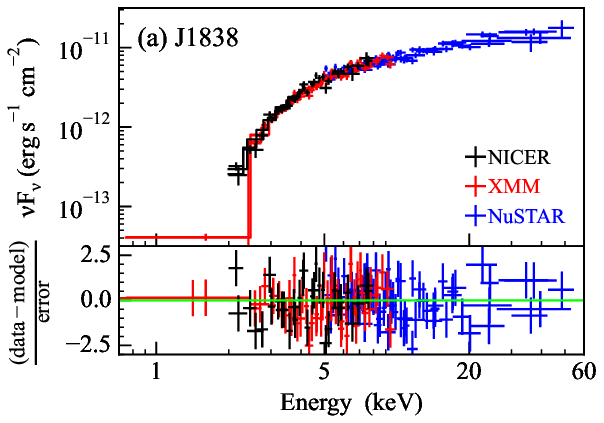}&
\hspace{-3 mm}
\includegraphics[width=2.3 in]{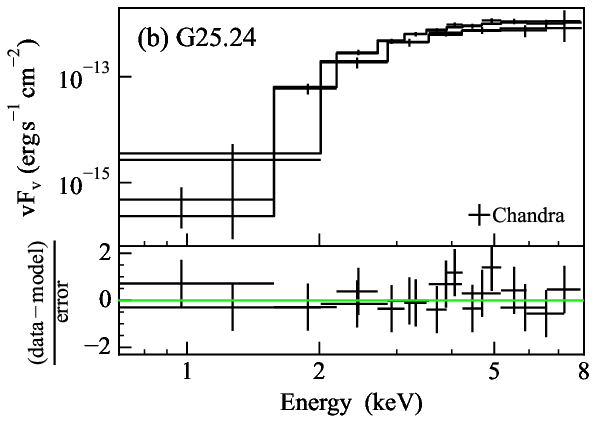}&
\hspace{-3 mm}
\includegraphics[width=2.3 in]{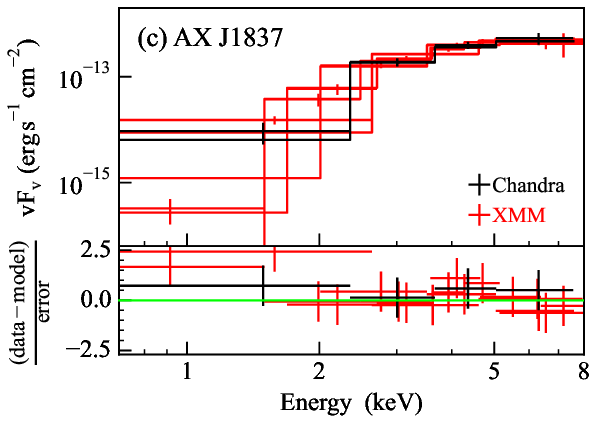}\\
\end{tabular}
\caption{Measured X-ray spectra and best-fit models for J1838 (a), G25.24 (b), and AX~J1837 (c). Data from various X-ray observatories are indicated by distinct colors. Solid lines represent the best-fit {\tt logpar} model for J1838 (a) and PL models for G25.24 (b) and AX~J1837 (c). Bottom panels in the figure display the residuals after subtracting the best-fit models.\label{fig:fig4}}
\end{figure*}

\subsection{Measurement of the X-ray Extent of G25.24}
\label{sec:sec2_3}
We determined the spatial extent of the PWN G25.24 using high-resolution Chandra data. While \citet{Gotthelf2008} identified the PWN emission extending to $\sim60''$ based on a 20\,ks Chandra observation (Obs. ID 6719), our analysis incorporates an additional 54\,ks of Chandra data (Obs. ID 16673) to more precisely measure the radial surface brightness profile. Although large off-axis angles of $5.5'$ and $7'$ for these observations complicate characterizing the PWN morphology, they do not significantly hinder the estimation of energy-dependent variations in the nebular size.

We extracted the radial profile from the combined Chandra data. The profile (Figure~\ref{fig:fig3}) reveals a sharp core attributable to the pulsar emission and clearly demonstrates that the PWN emission extends to $R\gapp 90''$ from the pulsar. Thanks to Chandra's high spatial resolution, we were able to estimate the PWN emission independent of the pulsar component by excluding the inner $10.5''$ region. The radial profile was modeled using an exponential function $ A\mathrm{exp}(-r/w)$ to represent the PWN, combined with a spatially constant background component. Instrumental vignetting was corrected for using exposure maps generated with the {\tt fluximage} tool.

We optimized the amplitudes of the emission components and the width parameter ($w$) of the exponential PWN model, yielding a best-fit value of $w=21.5''\pm 0.8''$ in the 1--8\,keV band. To investigate potential energy dependence, we performed an analysis of the data divided into two energy bands with approximately equal counts: 1--4\,keV (low) and 4--8\,keV (high), obtaining width values of $23.2''\pm 1.1''$ and $19.6''\pm1.0''$, respectively. Although this difference is not statistically significant at a high confidence level, the trend suggests a possible spectral softening with increasing radius, potentially due to synchrotron cooling of the PWN electrons.

\subsection{Spectral Analyses}
\label{sec:sec2_4}
We performed spectral analyses of the pulsar J1838, its associated PWN G25.24, and the PWN candidate AX~1837 (Figure~\ref{fig:fig1}) to verify the previously reported spectral curvature for J1838 \citep[][]{kh15,Takata2024} and to refine prior spectral measurements for the PWNe using larger Chandra and XMM-Newton datasets. For the measurement of the pulsar spectrum, an on$-$off analysis was applied to the high timing resolution data of XMM-Newton PN, NICER, and NuSTAR using our timing solutions to effectively remove emission from G25.24. Chandra's high spatial resolution minimized pulsar contamination in the spectral analysis of G25.24, while the high photon statistics of XMM-Newton improved the spectral analysis of AX~1837. 

Spectral fitting was conducted using XSPEC 12.14.0 with $\chi^2$ statistics. A cross-normalization factor was applied to each spectrum to account for potential cross-calibration issues between instruments \citep[e.g.,][]{Madsen2017}. Unless otherwise stated, Galactic absorption was modeled using \texttt{wilm} abundances \citep[][]{wam00} and \texttt{vern} cross-sections \citep[][]{vfky96}.

\subsubsection{On$-$off Spectrum of the Pulsar J1838}
\label{sec:sec2_4_1}
\begin{table*}[t]
\vspace{-0.0in}
\begin{center}
\caption{Spectral analysis results}
\label{ta:ta3}
\vspace{-0.05in}
\scriptsize{
\begin{tabular}{lcccccccccc} \hline\hline
Data       & {Instrument}\marka & Model\markb  & Energy & $N_{\rm H}$& $\Gamma$ & $E_{\rm b}$ & $\Gamma_2$  &  $\fluxtt$\markc& $\chi^2$/d.o.f.& Comment \\
           &   &     &(keV)   &($10^{22}\rm \, cm^{-2}$) &  & (keV) &  &  &   & \\ \hline
J1838 & XP+Ni+Nu & PL & 0.3--60 & $11.2\pm0.3$& $1.42\pm0.02$&$\cdots$ & $\cdots$  & $6.7\pm0.2$\markd& 605/576&  on$-$off \\ 
J1838 & XP+Ni+Nu & BPL & 0.3--60 & $8.9\pm 0.5$ & $1.02\pm0.08$ & $7.0\pm0.5$& $1.47\pm0.03$  & $6.3\pm0.2 $\markd& 569/574&  on$-$off \\
J1838 & XP+Ni+Nu & \texttt{logpar}& 0.3--60 & $9.0\pm0.5$ & $1.35\pm0.03$ & $10$  & $0.38\pm0.07$ & $6.3\pm0.2$\markd & 578/575 &  on$-$off \\
\hline
G25.24 & CXO& PL& 0.3--8& $9.1 \pm 1.2 $& $2.08 \pm 0.26$& $\cdots$& $\cdots$  & $2.04 \pm  0.13$ & 175/174 & $\cdots$ \\
\hline
AX~J1837 & CXO+XM & PL& 0.3--8& $8.2 \pm 1.2$ & $1.80 \pm 0.25$& $\cdots$& $\cdots$  & $0.81 \pm 0.09$& 287/280 & $\cdots$ \\
 \hline
\end{tabular}}
\end{center}
\vspace{-0.5 mm}
\footnotesize{
\marka{XP: XMM-Newton PN, XM: XMM-Newton MOS, Ni: NICER, Nu: NuSTAR, CXO: Chandra.}\\
\markb{Spectral models. PL: $K(E/1\, \mathrm{keV})^{-\Gamma}$. BPL: $K(E/1\, \mathrm{keV})^{-\Gamma}$ for $E\le E_{\rm b}$ and $K (E_{\rm b}/1\, \mathrm{keV})^{\Gamma_2 - \Gamma}(E/1\, \mathrm{keV})^{-\Gamma_2}$ for $E>E_{\rm b}$. {\tt logpar}: $K(E/E_{\rm b})^{[-\Gamma- \Gamma_2 \mathrm{log}(E/E_{\rm b})]}$.}\\
\markc{Absorption-corrected 2--10\,keV flux in units of $10^{-12}\,$\fluxcgs.}\\
\markd{Spin-cycle averaged flux.}\\
}
\end{table*}

For an on$-$off spectral analysis of XMM-Newton PN, NICER, and NuSTAR data, we extracted events within the on- and off-pulse intervals (Figure~\ref{fig:fig2}b) to construct source and background spectra, respectively. Spatial filtering was applied using circular regions of $R=16''$ for XMM-Newton and $R=60''$ for NuSTAR.
Spectral response files were generated using the standard tools: {\tt nuproduct} (NuSTAR), {\tt arfgen} and {\tt rmfgen} (XMM-Newton), and {\tt nicerl3-spect} (NICER).  The NICER monitoring data yielded 72 pre-glitch and 25 post-glitch spectra. These NICER spectra and corresponding response files within each of the pre- and post-glitch periods were merged using the \texttt{addascaspec} script.

We grouped these spectra to ensure a minimum of 200 counts per bin and fit jointly using a straight PL model and two curved PL models: BPL and log-parabolic \citep[\texttt{logpar};][]{Massaro2004} models.\footnote{https://heasarc.gsfc.nasa.gov/xanadu/xspec/models/logpar.html} While these models yielded statistically acceptable results with the best-fit parameters presented in Table~\ref{ta:ta3}, curved PL models were statistically preferred over the PL model, as indicated by F-test probabilities of $2.0\times10^{-8}$ and $3.3\times10^{-7}$ for the BPL and {\tt logpar} models, respectively. We show the {\tt logpar} model fit in Figure~\ref{fig:fig4}a for illustrative purposes. 

Based on these results, we estimated $E_{\rm SR}$. The best-fit parameters for the {\tt logpar} model imply an energy at the SED maximum of $E_{\rm SR}=73^{+85}_{-26}$\,keV. We verified that while different values of $E_{\rm b}$ altered the best-fit values of $\Gamma$ and $\Gamma_2$ reported in Table \ref{ta:ta3}, they did not affect the estimated $E_{\rm SR}$ value. 
Estimating $E_{\rm SR}$ with the BPL model is infeasible as it does not exhibit a spectral maximum. However, we investigated whether including a high-energy exponential cutoff ($e^{-E/E_c}$) in the BPL model would improve the fit. This modified model did not significantly improve the fit ($\chi^2$/d.o.f.=568/574), and it only provided a 68\% lower limit of $E_{\rm c}\ge72$\,keV, which translates to $E_{\rm SR}\ge 55$\,keV.

We note that our inferred $N_{\rm H}$ of $(9\textrm{--}11)\times{10^{22}}\,\mathrm{cm}^{-2}$ is significantly higher than $(5\textrm{--}7)\times 10^{22}\rm \, cm^{-2}$ reported for the pulsar and its PWN in previous studies \citep[e.g.,][]{Kargaltsev2012,Takata2024}. We attribute this discrepancy primarily to differences in adopted abundance tables. While these previous works did not specify their chosen abundance table, we suspect that they used the default \texttt{angr} table in XSPEC, given the consistency of their $N_{\rm H}$ values with that of \citet{Anada2009} who used the \texttt{angr} table. Indeed, using the \texttt{angr} abundances in our analysis, we obtained $N_{\rm H}=(6\textrm{--}7) \times 10^{22}\rm \, cm^{-2}$, consistent with the previous reports.

\subsubsection{X-ray Spectrum of the PWN G25.24}
\label{sec:sec2_4_2} 
\cite{Gotthelf2008} analyzed a 20-ks Chandra observation (Obs. ID 6719) and measured G25.24's spectrum within an $R=60''$ circle excising the central $5'' \times 7''$ to remove the pulsar contribution. They found that the spectrum is well described by a PL with $\Gamma=1.6\pm0.4$ and the 2--10\,keV flux of $\fluxtt=1.0 \times 10^{-12}$\,\fluxcgs\ for $N_{\rm H}=4.5^{+0.7}_{-0.8}\times 10^{22}\,\mathrm{cm}^{-2}$ (90\% confidence interval). However, we measured the PWN extension to be $R\gapp90''$ (Figures~\ref{fig:fig1} and ~\ref{fig:fig3}), substantially larger than the $R=60''$ region used by \citet{Gotthelf2008}. This necessitates a revision to the previous spectral results.

To measure the spectrum of G25.24, we analyzed 74\,ks of Chandra data (Obs. IDs 6719 and 16673). The pulsar emission was excised using elliptical regions with radii of $10.5''\times7.5''$. While the position angles of these elliptical excision regions differed between the two observations due to varying off-axis positions, this difference is negligible compared to the larger emission region of G25.24.
Source spectra were extracted from a circular region of radius $R=90''$ (Figure~\ref{fig:fig1}), centered on the pulsar position. Background spectra were obtained from nearby source-free regions. We fit the source spectra in the 0.3--8\,keV band with an absorbed PL model (Figure~\ref{fig:fig4}b). The best-fit parameters are presented in Table~\ref{ta:ta3}. We note that the bright central region of G25.24 was located on a chip gap in the earlier observation (Obs. ID 6719). This resulted in a reduced measured flux compared to the later observation, which we accounted for with a cross-normalization constant in the fit.

Although statistically consistent, our derived $\Gamma$ is slightly softer and the flux is higher than the previous measurements of $\Gamma=1.6$ and $\fluxtt=1.0\times 10^{-12}$\,\fluxcgs. This difference in $\Gamma$ was found to be attributable to the different $N_{\rm H}$ values employed.  Using the $N_{\rm H}$ value of $4.5\times 10^{22}\rm \ cm^{-2}$ and adopting the {\tt angr} abundance table, we obtained $\Gamma=1.6\pm 0.2$.
The flux discrepancy, however, arises from a combination of factors: 
(1) variations in background region selections ($\sim 5$\%), 
(2) discrepancy in $N_{\rm H}$ values ($\sim 10$\%), 
(3) differences in source region sizes ($\sim 25$\%), and
(4) a flux loss due to the chip gap in Obs. ID 6719 ($\sim 30$\%).

\subsubsection{Characterization of the X-ray Emission from the Candidate PWN AX~J1837}
\label{sec:sec2_4_3} 
The candidate PWN AX~J1837 is located $\sim10'$ west of PSR~J1838$-$0655 (Figure~\ref{fig:fig1}) and was reported as an extended X-ray source by \citet{Gotthelf2008}, based on Chandra data (Obs. ID 6719). Their analysis resolved AX~J1837 into a faint point source surrounded by a diffuse nebula with the nebular flux dominating over the point source component. The spectral parameters for the nebular emission
were largely unconstrained, with only rough estimates of $\Gamma$ and  $N_{\rm H}$ of  $0.7\textrm{--}3.6$ and $(2\textrm{--}12)\times 10^{22}\,\mathrm{cm}^{-2}$ (90\% confidence intervals), respectively \citep[see also][]{Kargaltsev2012}.

We assessed the contribution of the faint point source using the Chandra data (Obs. ID 6719). The net 0.3--8\,keV count ratio between the point source and the extended emission was estimated to be approximately 3--4\%, indicating that the point-source emission is negligible. A spectral fit of the point-source data with a PL model further showed that its flux is $\sim6$\% of the total flux. We also verified that the point source does not significantly impact $\Gamma$ of the extended emission by performing spectral analysis with and without its inclusion. Therefore, we included the point source in our analysis and performed a joint analysis of one Chandra (Obs. ID 6719) and two XMM-Newton observations (Table~\ref{ta:ta1}). The source was outside the field of view in the longer Chandra  (Obs. ID 16673) and the NuSTAR observations. 

The spectra of AX~J1837 were extracted from elliptical regions of $60''\times 90''$ (Figure~\ref{fig:fig1}), and background spectra were obtained from nearby source-free regions.
Joint spectral fitting of all three observations was performed in the 0.3--8\,keV band using an absorbed
PL model (Figure~\ref{fig:fig4}c). The best-fit spectral parameters obtained from our analysis are presented in Table~\ref{ta:ta3}.

The inclusion of the XMM-Newton data yielded more precise constraints on $N_{\rm H}$ and $\Gamma$ compared to the previous Chandra results. Our derived $N_{\rm H}$ value is consistent with that of J1838, suggesting a similar distance for AX~J1837. The $\Gamma=1.8\pm0.3$ falls within the typical range for PWNe. We note that this source was observed at large off-axis angles ($5'$ for Chandra and $11'$ for XMM-Newton), near the edge of the observatories' field of view. This resulted in very low collected counts and, consequently, large statistical uncertainties. Furthermore, a noticeable covariance between $N_{\rm H}$ and $\Gamma$ was present, leading to substantial variations (comparable to the statistical uncertainties) in their inferred values depending on the background selection.

\section{SED modeling}
\label{sec:sec3}
\begin{figure*}[t]
\centering
\begin{tabular}{cc}
\vspace{0 mm}
\includegraphics[width=3.3 in]{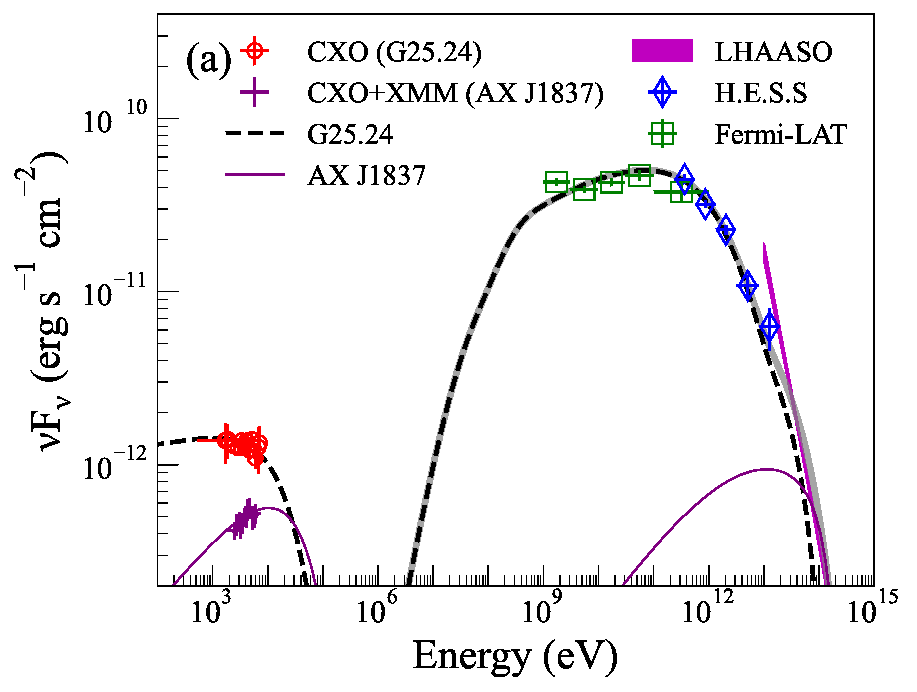} &
\raisebox{-0.  mm}{\includegraphics[width=3.3
in]{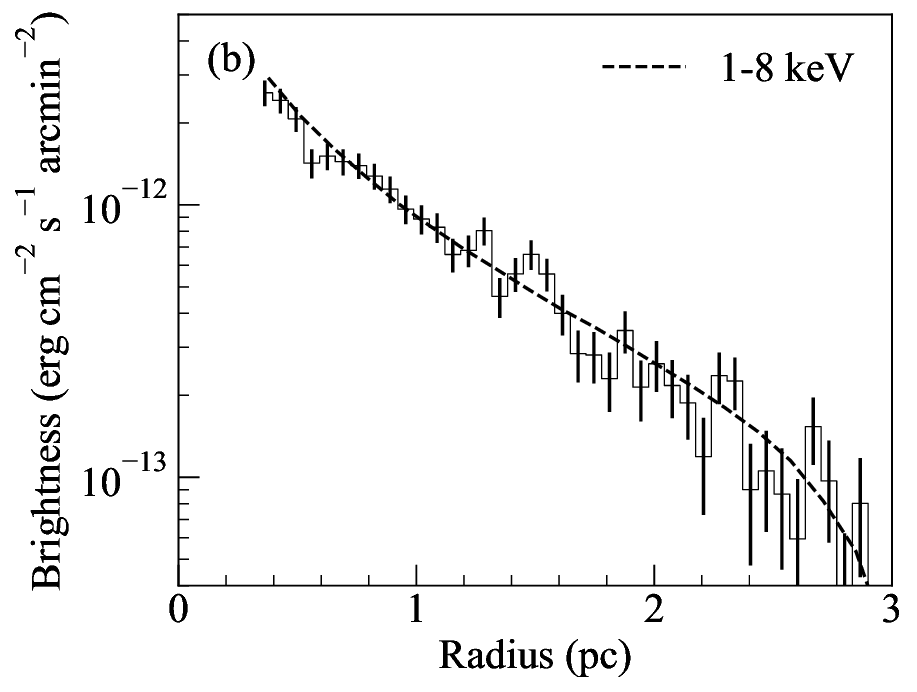}} \\
\end{tabular}
\caption{(a) Broadband SED of the emission from the region around HESS~J1837$-$069 (e.g., Figure~\ref{fig:fig1}) measured by Chandra and XMM-Newton (red and purple for G25.24 and AX~J1837, respectively; Section~\ref{sec:sec2_4}), Fermi LAT \citep[green squares;][]{Ballet2023}, H.E.S.S. \citep[blue diamonds;][]{HESSHGPS2018}, and LHAASO \citep[magenta line;][]{Cao2024}. SED models for G25.24 and AX~J1837 are shown in black dashed and purple solid curves, respectively. The thick gray curve represents the combined VHE SED model. (b) Background-subtracted radial profile of the X-ray surface brightness of G25.24. We converted counts from Figure~\ref{fig:fig3} into flux units by normalizing them with the measured flux in Table~\ref{ta:ta3}. The dashed curve shows our PWN model computation.}
\label{fig:fig5}
\vspace{0mm}
\end{figure*}

To investigate the VHE and UHE emissions from the HESS~J1837$-$069 region, we constructed a broadband SED. This SED incorporated our X-ray measurements (Sections~\ref{sec:sec2_4_2} and \ref{sec:sec2_4_3}) along with published Fermi-LAT \citep{Ballet2023},\footnote{\href{https://fermi.gsfc.nasa.gov/ssc/data/access/lat/14yr_catalog/gll_psc_v35.fit}{https://fermi.gsfc.nasa.gov/ssc/data/access/lat/14yr\_catalog/\\gll\_psc\_v35.fit}} H.E.S.S. \citep{HESSHGPS2018},\footnote{\href{https://www.mpi-hd.mpg.de/hfm/HESS/hgps/data/hgps_catalog_v1.fits.gz}{https://www.mpi-hd.mpg.de/hfm/HESS/hgps/data/hgps\_\\catalog\_v1.fits.gz}} and LHAASO \citep{LHAASO2023} data (Figure~\ref{fig:fig5}a).

The region surrounding HESS~J1837$-$069 exhibits significant complexity (Figure~\ref{fig:fig1}). The H.E.S.S. data necessitates three distinct components \citep[][]{HESSHGPS2018}: a compact region ($R=0.07^\circ$) partially overlapping G25.24, an extended region encompassing both G25.24 and AX~J1837 (as indicated in Figure~\ref{fig:fig1}), and a faint region located $0.5^\circ$ south of J1838. This latter component contributes only $\approx4$\% of the total H.E.S.S. flux. Similarly, the Fermi-LAT data \citep[][]{Ballet2023} require two components \citep[see also][]{Katsuta2017, Sun2020}.

Given the spatial overlap between G25.24 and the compact component of HESS~J1837$-$069, the PWN G25.24 is likely a significant contributor to the VHE emission. Additionally, the positional coincidence between AX~J1837 and 1LHAASO~J1837$-$0654u (Figure~\ref{fig:fig1}) strongly suggests that AX~J1837 contributes significantly to the observed UHE emission. 

Consequently, we propose a leptonic scenario in which the broadband emission from the HESS J1837$-$069 region originates from two PWNe: G25.24 and AX J1837. While the exact association of the two Fermi-LAT sources with HESS~J1837$-$069 remains unclear \citep[e.g.,][]{Katsuta2017,Sun2020}, we assume their combined emission arises from these PWNe. We then test the feasibility of this scenario using a PWN emission model. Alternative scenarios are discussed in Section \ref{sec:sec4_3}.

The SED model, which incorporates advection and diffusion processes, and adiabatic and radiative losses, has been previously applied to other middle-aged PWNe \citep[see][for detailed descriptions and applications]{Park2023a,Park2023b}. A brief overview of the model is presented in Section~\ref{sec:sec3_1}.

\subsection{Model Description}
\label{sec:sec3_1}

We assume that the pulsar injects electrons at the termination shock with a PL distribution between $\gamma_{e,\rm min}$ and $\gamma_{e,\rm max}$: 
\begin{equation}
\label{eq:edist}
\frac{dN_e}{d\gamma_e dt}=N_0\,\gamma_e^{-p_1},
\end{equation}
where $\gamma_e$ is the Lorentz factor of the electrons. 
The temporal evolution of the pulsar's $\dot E_{\rm SD}(t)$ is modeled by
\begin{equation}
\label{eq:Edot}
\dot E_{\rm SD}(t)=\dot E_{\rm SD}(0) \left ( 1 + \frac{t}{\tau_0} \right )^{-\frac{n+1}{n-1}},
\end{equation}
where $\tau_0=2\tau_c/(n-1) - t_{\rm age}$ \citep[e.g.,][]{Gelfand2015}. Here, $\tau_c$ is the pulsar's characteristic age, and $t_{\rm age}$ is the assumed PWN age. We adopt a braking index of $n=3$.

We assume that the energy injected by the pulsar is partitioned between particles and the magnetic field within the PWN with fractions of $\eta_e$ and $\eta_B$, respectively, such that $\eta_e + \eta_B=1$.
The particle injection rate evolves with time: 
\begin{equation}
\label{eq:injecte}
\int_{\gamma_{e,\rm min}}^{\gamma_{e,\rm max}} \gamma_e m_e c^2 \frac{dN_e}{d\gamma_e dt} d\gamma_e=\eta_e \dot E_{\rm SD}(t).
\end{equation}
This time-dependent injection implies that the pulsar supplied more energy and particles to the PWN in its early stages. These older electrons subsequently cool and propagate outward over longer periods, contributing significantly to the VHE emission observed beyond the X-ray PWN radius ($R_{\rm PWN}$).

The following PL prescriptions are adopted for the PWN properties, as in \citet{Park2023b}:
\begin{equation}
\label{eq:vflow}
V_{\rm flow}(r)= \begin{cases}
 V_0 \left( \frac{r}{R_{\rm TS}} \right )^{\alpha_V}, & \textrm{for } R_{\rm TS} \le r \le R_{\rm PWN} \\
 V_{\rm ext}, & \textrm{for } R_{\rm PWN} < r \le R_{\rm max} \\
\end{cases}
\end{equation}
for the flow speed (governing advection and adiabatic losses),
\begin{equation}
\label{eq:bfield}
B(r)= \begin{cases}
 B_0 \left( \frac{r}{R_{\rm TS}} \right )^{\alpha_B}, & \textrm{for } R_{\rm TS} \le r \le R_{\rm PWN} \\
 B_{\rm ext}, & \textrm{for } R_{\rm PWN} < r \le R_{\rm max}  \\
\end{cases}
\end{equation}
for the magnetic field $B$ (determining synchrotron radiation), and
\begin{equation}
\label{eq:diff}
D(r,\gamma_e)=D_0 \left( \frac{B(r)}{100\mu\rm G} \right )^{-1} \left( \frac{\gamma_e}{10^9} \right )^{1/3}
\end{equation}
for the diffusion coefficient.  Here, $R_{\rm TS}$ and $R_{\rm max}$ denote the radii of the termination shock and our computation volume, respectively. We assume a toroidal $B$ structure \citep[e.g.,][]{CrabIXPE2023} and magnetic flux conservation within the PWNe, which implies the relation $\alpha_V + \alpha_B=-1$. 

Within the X-ray PWN radius, the magnetic field follows this PL trend to account for the observed X-ray flux and brightness profile (Figure~\ref{fig:fig5}b) while satisfying the energy condition:
\begin{equation}
\label{benergy}
\int_{R_{\rm TS}}^{R_{\rm PWN}} \frac{B(r)^2}{8 \pi} 4\pi r^2 dr=\eta_B \int_{0}^{t_{\rm age}} \dot E_{\rm SD}(t) dt.
\end{equation}
Beyond this radius, we adopt constant interstellar values for the external magnetic field ($B_{\rm ext}$) and external flow speed ($V_{\rm ext}$).

The injected electrons propagate outwards, undergoing energy losses. This process is simulated by stepping forward in time. At each time step, electron injection is computed using Equations~(\ref{eq:edist})--(\ref{eq:injecte}). Concurrently, advective transport is calculated using Equation~(\ref{eq:vflow}), and diffusive transport is simulated using the diffusion coefficient given by Equation~(\ref{eq:diff}). Adiabatic losses are accounted for using the flow speed (Equation~\ref{eq:vflow}), while synchrotron and IC radiation are computed based on $B$ (Equation~\ref{eq:bfield}), the CMB, and an assumed IR photon field with energy density $U_{\rm IR}$ and blackbody temperature $T_{\rm IR}$. This procedure is iterated over the PWN's age.

We apply this procedure to both G25.24 and AX J1837. We compute the evolution of electrons and their subsequent synchrotron and IC emissions within a large volume extending to a radius of $R_{\rm max}$. For each source, these emissions are projected onto the tangent plane of the observer to construct the broadband SED and radial profile (Figure~\ref{fig:fig5}). For the spatially-integrated SED, the synchrotron emission is integrated within a projected radius of $R_{\rm pwn}$, while the IC emission is integrated within 40\,pc. These projected radii were selected to match the observed X-ray and H.E.S.S. angular sizes of the respective sources (Figure~\ref{fig:fig1}). We then combine their resulting emissions for comparison with the observed SEDs.

\subsection{Initial Consideration for the Model Parameters}
\label{sec:sec3_2}
The PWN G25.24 exhibits significantly brighter and slightly softer X-ray emission compared to AX~J1837, suggesting a dominant contribution to the observed VHE emission. This allows us to estimate initial model parameters for G25.24 using a single-zone approach, based on the observed X-ray and VHE properties of G25.24 and HESS~J1837$-$069, respectively.

\begin{table*}[t]
\vspace{-0.0in}
\begin{center}
\caption{Parameters for the SED model in Figure~\ref{fig:fig5}}
\label{ta:ta4}
\vspace{-0.05in}
\scriptsize{
\begin{tabular}{lcccc} \hline\hline
Parameter  & Symbol  &  Unit  & G25.24  & AX~J1837   \\ \hline
Spin-down power       & $\dot E_{\rm SD}$  &  $10^{36}\rm \ erg\ s^{-1}$  &  5.5  & 0.7    \\
Age of the PWN        & $t_{\rm age}$  &  kyr  & 21.5 &  19   \\
Size of the PWN       & $R_{\rm pwn}$  & pc & 3 & 3      \\
Radius of termination shock & $R_{\rm TS}$ & pc & 0.1 & 0.1  \\
 Radius of computation volume & $R_{\rm max}$ & pc & 100 & 100  \\
Distance to the PWN  & $d$  &   kpc  & 6.6  & 6.6  \\ 
Index for the particle distribution   & $p_1$  &  & 2.5   & 1.9          \\
Minimum Lorentz factor  & $\gamma_{e,\rm min}$ &  & $10^{5.2}$  & $10^{3.0}$ \\
Maximum Lorentz factor  & $\gamma_{e,\rm max}$ &  & $10^{8.6}$  & $10^{8.8}$   \\
Magnetic field        & $B_0$   & $\mu$G   &  37.9  & 10.7   \\
Magnetic index        & $\alpha_B$ &  & $-0.4$ & $-0.05$ \\
 External magnetic field & $B_{\rm ext}$  & $\mu$G   &  3  & 3   \\
Flow speed            & $V_0$  &  $c$  & 0.01   & 0.005 \\
Speed index           & $\alpha_V$  &  & $-0.6$  & $-0.95$ \\
 External flow speed & $V_{\rm ext}$  & \kms\   &  0  & 0   \\
Diffusion coefficient & $D_0$  & $10^{27} \,\rm \ cm^2 \ s^{-1}$ & 1.2 & 0.5 \\
Energy fraction injected into particles   &  $\eta_e$  &   &  0.99  &  0.99  \\
Energy fraction injected into $B$  & $\eta_B$  &  & 0.0002  & 0.006    \\
Temperature of IR seeds & $T_{\rm IR}$  & K & 20 & 20 \\
Energy density of IR seeds & $U_{\rm IR}$ & $\,\mathrm{eV}\,\mathrm{cm}^{-3}$ & 1.1 & 1.1 \\ \hline
\end{tabular}}
\end{center}
\footnotesize{Note. In the absence of a detected pulsar, $\dot E_{\rm SD}$ for AX~J1837 was assumed.}
\vspace{-0.5 mm}
\end{table*}

The observed X-ray-to-TeV flux ratio of $\sim 0.05$ \citep[Table~\ref{ta:ta3} and][]{HESSHGPS2018} allows an estimation of the PWN $B$ through the relationship $F_{\rm X}/F_{\gamma} = U_B/U_{\rm IR}$ (ignoring the Klein-Nishina effect) provided that both emissions originate from the same electron population (i.e., one-zone approach). Assuming $U_{\rm IR}=1$\,\evcm\ at $T_{\rm IR}=20$\,K,  we estimate $B$ within G25.24 of $B \approx 1.5\,\mu$G. Since this is too low (e.g., lower than intergalactic $B$ of $\sim2\textrm{--}3\,\mu$G), we initially assume $B=3\,\mu$G. The flux ratio also suggests G25.24 age of $\gapp10$\,kyr \citep[e.g.,][]{zhu2018}. We assume an age of $t_{\rm age}\sim 20$\,kyr, comparable to the pulsar's $\tau_c$ of 22.5\,kyr.

Given that the X-ray emission arises from radiatively cooled electrons and that synchrotron cooling induces a spectral softening of $\Delta \Gamma\sim 0.5$ above a break energy, the measured $\Gamma = 2.08$ for G25.24 suggests that the emission from freshly accelerated (uncooled) electrons would exhibit $\Gamma\approx 1.6$. This corresponds to an electron spectral index of $p_1\approx 2.2$ (Equation~\ref{eq:edist}). The X-ray emission implies $\gamma_e \gapp 10^8$ for the assumed $B=3\,\mu$G. The parameters $\gamma_{e,\rm min}$, $\eta_e$, and $\eta_B$ were adjusted to conserve the energy supplied by the pulsar.

Observations of UHE emission coincident with AX~J1837 by LHAASO enable an estimation of its maximum electron Lorentz factor, $\gamma_{e,\rm max}$. The energy of IC-upscattered photons, $E_{\gamma, \rm IC}$, can be approximated by $E_{\gamma, \rm IC}\approx (4/3) \gamma_e^2 E_{\gamma, \rm seed}$, where $E_{\gamma, \rm seed}$ is the seed photon energy. Given that the UHE emission predominantly arises from IC scattering of CMB photons (due to Klein-Nishina suppression for IC scattering of IR seed photons), the detection of $\gapp 100$\,TeV emission by LHAASO (Figure~\ref{fig:fig5}a) suggests $\gamma_{e,\rm max}\gapp 10^{8.5}$. It is important to note that this is a rough initial estimate, as it does not include the effects of particle cooling, the Klein-Nishina cross-section, or the shape of the SED. Our model incorporates these crucial physical processes and fits the SED shape, thereby providing a more robust estimation of $\gamma_{e,\rm max}$ (Section~\ref{sec:sec3_3}). In particular, distinct features (e.g., cutoffs) observed in the X-ray and/or VHE SEDs, are essential for precisely constraining $\gamma_{e,\rm max}$.

The termination shock of this PWN has not been identified in the X-ray images. So we adopt $R_{\rm TS}=0.1$\,pc as measured for other PWNe \citep[][]{Kargaltsev2008}. The exact $R_{\rm TS}$ value does not have large impact on the model results. The observed X-ray size of $R=90''$ corresponds to $R_{\rm PWN}=3$\,pc. A case exploring a larger X-ray extension is presented in Section~\ref{sec:sec3_3}. The VHE size of the PWN (e.g., HESS~J1837$-$069) provides an estimate for the diffusion coefficient. Given that VHE emission arises from IC scattering by electrons with $\gamma_e\sim 10^7$, the measured size of $\sim40$\,pc gives $D_0 \sim 10^{27}\rm \ cm^{-2}$. For this $D_0$ value, the highest-energy electrons can diffuse out to $\sim 100$\,pc. Consequently, we adopt $R_{\rm max}=100$\,pc.

We adopt similar initial parameters for AX~J1837. Because its pulsar has not been identified, we assumed a lower $\dot E_{\rm SD}$ and a similar $t_{\rm age}$ based on its fainter emission and comparable X-ray extension with G25.24, respectively.
These initial values, while serving as a starting point, are significantly refined through the application of our model in Section~\ref{sec:sec3_3}.

\subsection{Modeling of Emission from HESS~J1837$-$069}
\label{sec:sec3_3}

Although the initial parameter set provided a preliminary SED representation, it does not capture the complexity of energy-loss processes and particle evolution. Consequently, significant parameter revisions were necessary. While the measured spin properties of PSR~J1838$-$0655 permit the modeling of emission from G25.24, detailed SED modeling of AX~J1837's emission is hampered by the lack of detected pulsations, which are crucial for constraining the energy injection rate (i.e., $\dot E_{\rm SD}$) {\bf and age ($t_{\rm age}$)} of the PWN. Therefore, the model parameters for AX~J1837 are not well constrained. Nonetheless, for our SED study, we posited that AX~J1837 is a PWN powered by a middle-aged pulsar \citep[e.g.,][]{Gotthelf2008,Kargaltsev2012} and that its emission dominates in the UHE band,  a scenario supported by the positional coincidence between AX~J1837 and the LHAASO source (Figure~\ref{fig:fig1}).

Since the observed VHE emission likely originates primarily from G25.24, modeling the VHE data enables estimations of its physical properties. The measured radial surface brightness profile of G25.24 in the X-ray band and the angular extent of the H.E.S.S. region provide constraints on its advection ($V_0$, $\alpha_V$, and $\alpha_B$) and diffusion ($D_0$) parameters, respectively. Matching the observed UHE SED at $\gapp 10^{14}$\,eV with emission from AX~J1837 allows estimations of its parameters such as $\gamma_{e, \rm max}$, as explained in Section~\ref{sec:sec3_2}.
To reproduce the observed broadband SED (Figure~\ref{fig:fig5}a) and the shape of the X-ray surface brightness profile (Figure~\ref{fig:fig5}b) with our multi-zone model, we iteratively refined the initial model parameters through visual inspection. The refined model and its parameter values are presented in Figure~\ref{fig:fig5} and Table~\ref{ta:ta4}.

Some of the derived parameter values differ substantially from the initial assumptions presented in Section~\ref{sec:sec3_2}. Specifically, our SED modeling necessitated a significantly higher $B$ for G25.24 than initially estimated. This increase is required to reconcile the observed high X-ray flux with the pulsar's $\dot E_{\rm SD}$. For the derived $B$ values, our model predicts strong spectral curvatures at energies $\gapp 10$\,keV for both G25.24 and AX~J1837 due to high-energy cutoffs in their synchrotron emissions \citep[e.g.,][]{An2019}. Deep NuSTAR observations could confirm these predictions.

The required IR seed photon density of 1\,\evcm\ (constrained by properties of G25.24 and the VHE emission) is comparable to the Galactic average. This value, however, is substantially lower than that implied by the $F_{\rm X}/F_{\gamma} = U_B/U_{\rm IR}$ relationship. This discrepancy arises because the VHE emission originates from a larger region extending beyond the X-ray PWNe, rendering the one-zone approach invalid for this source. In our model, we can roughly estimate the spatial extent of the VHE emission from HESS J1837$-$069. For $D_0 = 1.2 \times 10^{27}$ cm$^2$ s$^{-1}$ and $B_{\rm ext}=3\mu$G, the diffusion coefficient for $\gamma_e \sim 10^7$ electrons (which are responsible for IC emission in the VHE band) is estimated to be $9\times 10^{27}\rm \ cm^{2}\ s^{-1}$. This corresponds to a diffusion length scale of $2\sqrt{Dt_{\rm age}} \sim 50$\,pc, which is significantly larger than the X-ray size of G25.24 ($\approx 3$\,pc) and similar to the TeV size of HESS J1837$-$069.

This means that electrons injected when the PWN was young contribute to VHE emission through IC scattering outside the synchrotron emission region (i.e., X-ray PWN), whereas X-ray emission is dominated by fresh, energetic electrons within that region. Therefore, middle-aged or evolved PWNe often exhibit stronger (and more extended) VHE emission relative to X-ray emission.

For this reason, the parameters $t_{\rm age}$ and $U_{\rm IR}$ covary. Younger PWN ages necessitate higher $U_{\rm IR}$ values. Assuming that the IR background around J1838 is likely comparable to the Galactic average of a few \evcm, models with young PWN ages appear less plausible. Conversely, larger values of $t_{\rm age}$ enhance electron cooling, leading to a reduction in the VHE flux at the highest energies. Our analysis indicates that ages of $\sim20$\,kyr for the two PWNe adequately reproduce the observed VHE SED, despite the $t_{\rm age}$ (and other model parameter) value for AX~J1837 being highly uncertain due to the lack of an identified pulsar.

While the parameter values presented in Table~\ref{ta:ta4} can accommodate the observational data, they can vary substantially depending on the model assumptions. To assess the robustness of our results, we conducted additional tests utilizing model parameters that are different from our default selections (e.g., $B_{\rm ext}$, $V_{\rm ext}$, $R_{\rm PWN}$, and $n$).

We investigated variations in the PWN properties (Equations~\ref{eq:vflow} and \ref{eq:bfield}), including different values for $B_{\rm ext}$ and $V_{\rm ext}$, and scenarios where the PL trends for $V_{\rm flow}(r)$ and $B(r)$ extend beyond $R_{\rm PWN}$. This latter case probes the possibility that the X-ray PWN size is actually larger than observed (i.e., $R_{\rm PWN}>3$\,pc; undetected faint emission at large radii). While the influence of $V_{\rm ext}$ on the emission SED is negligible, $B_{\rm ext}$ affects the model predictions due to its role in particle cooling. However, the changes caused by variations in $B_{\rm ext}$ (e.g., $B_{\rm ext}=2\textrm{--}4\mu$G or extending the PL trend) are not substantial, as energy losses at low $B$ are minor. We found that the model-computed SED changes by $\lapp30$\%, which can be readily accommodated by small adjustments (e.g., $\lapp20$\%) of other parameters like $B_0$, $U_{\rm IR}$, and $D_0$.

The energy injection history of the pulsar, characterized by the braking index $n$ (Equation~\ref{eq:Edot}), also influences the SED. Although we assumed a value for magnetic braking, the true braking index for J1838 is unknown. To explore other possibilities, we tested a case with $n=2.5$, which resulted in a reduction in the broadband emission from the PWN due to a lower energy injection rate. This reduction could, however, be compensated for by adjusting other parameters such as $\gamma_{e,\rm min}$, $B_0$, $p_1$, and $U_{\rm IR}$.

We estimated the maximum electron energy within the PWNe to be $\sim$300\,TeV, corresponding to $\gamma_{e,\rm max}=10^{8.8}$ for AX~J1837. Our modeling indicates that altering this maximum energy by a factor of two causes the predicted SED to deviate significantly from the LHAASO-measured SED. 
However, this assessment was performed via visual inspection of SED data that lacked notable features, rather than a formal fit. Such an approach can lead to considerable variations in $\gamma_{e,\rm max}$, as demonstrated by differences in values inferred from SED modeling of nearly identical datasets \citep[e.g.,][]{Torres2014,zhu2018}. Moreover, uncertainties in the PL representation of the UHE SED and potential cross-calibration issues among gamma-ray instruments preclude an accurate determination of the maximum electron energy. Therefore, our reported $\gamma_{e,\rm max}$ value should be considered approximate.

In our current model, AX~J1837's contribution to the gamma-ray SED is minor, attributed to its fainter X-ray emission and harder X-ray spectrum. However, given the significant uncertainties in the X-ray spectra (Section~\ref{sec:sec2_4_3}) and the statistical consistency of $\Gamma$ for the two PWNe, AX~J1837's gamma-ray flux could potentially increase to $\sim40$\% of G25.24's, as suggested by their respective X-ray fluxes. These uncertainties are compounded by the fact that the nature of AX J1837's energy source (e.g., a pulsar) is unknown, which leaves parameters such as $\dot E_{\rm SD}$ and $t_{\rm age}$ unconstrained. Since the other parameter values can also vary depending on the model assumptions, the values presented in Table~\ref{ta:ta4} are highly covariant and uncertain. Moreover, they were derived through visual inspection rather than a formal data fit. Therefore, these derived values should be regarded as demonstrating the feasibility of the two-PWN scenario \citep[e.g., comparable to values found in other PWNe;][]{Torres2014,zhu2018} rather than as precise measurements.

This particular scenario, invoking emission from two PWNe, represents one plausible interpretation of the observed broadband emission characteristics of HESS~J1837$-$069; alternative possibilities are discussed in Section~\ref{sec:sec4_3}.

\section{Discussion}
\label{sec:sec4}
We investigated the rotational and emission properties of the MeV pulsar J1838 using X-ray observations from XMM-Newton, NICER, and NuSTAR. Our timing analysis yielded timing solutions which enabled a precise measurement of the pulsar's X-ray spectrum. The spectrum is well described by a BPL or {\tt logpar} model, thus confirming the presence of spectral curvature previously suggested in the literature.

We presented improved spectral characterizations of G25.24 and AX~J1837 by utilizing more extensive X-ray datasets.
By combining these new X-ray results with published gamma-ray measurements, we proposed G25.24 and AX~J1837 to be the primary contributors to the observed VHE and UHE emissions from the region, respectively, and demonstrated the feasibility of this scenario through SED modeling.

\subsection{The Pulsar J1838}
\label{sec:sec4_1}
As demonstrated in Section~\ref{sec:sec2_4_1}, J1838's X-ray spectrum exhibits curvature.
Using the best-fit {\tt logpar} parameters (Table~\ref{ta:ta3}), we estimated $E_{\rm SR}=73^{+85}_{-26}$\,keV. However, the BPL model with an exponential cutoff yields $E_{\rm SR} \geq 55$\,keV. This model dependence underscores that estimations of $E_{\rm SR}$ (for J1838 and other MeV pulsars) are subject to considerable uncertainties.

In this work, we adopt the $E_{\rm SR}$ value estimated using the {\tt logpar} model. This selection not only minimizes model-induced bias by aligning with prior $E_{\rm SR}$ estimations for other MeV pulsars \citep[e.g., PSR~B1509$-$58, PSR~J1846$-$0258, and PSR~J1849$-$0001;][]{Chen2016, Kuiper2018, Kim2024}, but also benefits from its theoretical foundation. The {\tt logpar} model has been suggested to adequately represent the energy of the synchrotron SED peak \citep[][]{Massaro2004}.

\begin{figure}[t]
\centering
\hspace{0.7 mm}
\includegraphics[width=3 in]{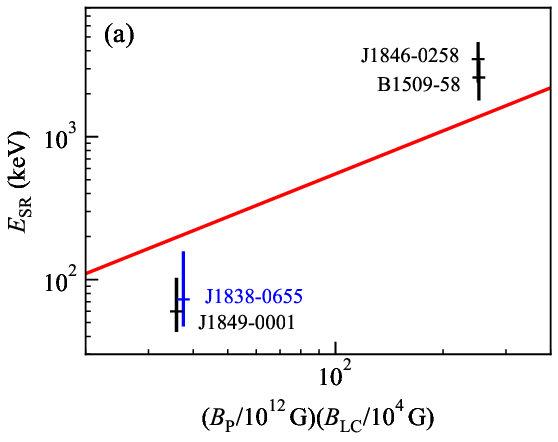}
\hspace{-5mm}
\includegraphics[width=3.065 in]{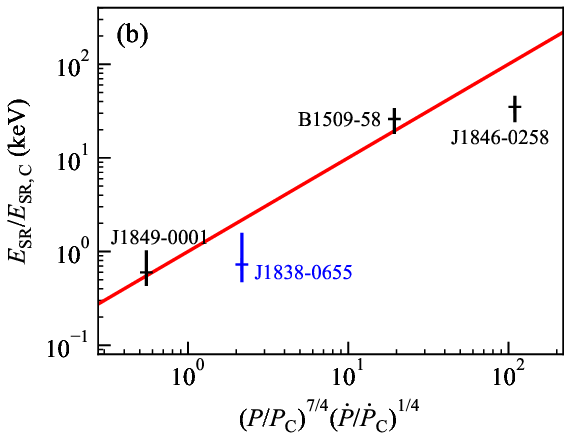}
\caption{Measured synchrotron peak energies ($E_{\rm SR}$) of four MeV pulsars \citep[Section~\ref{sec:sec2_4_1} and][]{Kim2024} plotted against combinations of magnetic-field strengths (at the surface $B_{\rm P}$ and light cylinder $B_{\rm LC}$; a) or spin parameters (b). The red lines represent the analytic relations of $E_{\rm SR} \propto B_{\rm P} B_{\rm LC}$ \citep[a;][]{Harding2017} and $E_{\rm SR} \propto P^{7/4} \dot{P}^{1/4}$ \citep[b;][]{Cusumano2001}.
\label{fig:fig6}}
\vspace{0mm}
\end{figure}

We incorporate our $E_{\rm SR}$ estimation for J1838 into the existing MeV pulsar dataset from \citet{Kim2024}, along with the relationship between $E_{\rm SR}$ and magnetic-field strengths (Figure~\ref{fig:fig6}a), derived by \citet{Harding2017} under the assumption that pairs are produced near the neutron star surface via magnetic pair production and subsequently emit synchrotron radiation near the light cylinder.  While the data generally follow this trend, they exhibit a slight deviation from the proposed relation (red line in Figure~\ref{fig:fig6}a).

The measurements also appear consistent with the relation $(E_{\rm SR}/E_{\rm SR,\, C})\approx (P/P_{\rm C})^{7/4} (\dot{P} / \dot{P}_{\rm C})^{1/4}$  (Figure~\ref{fig:fig6}b) proposed by \citet{Cusumano2001}, where the subscript ``C'' represents the values for the Crab pulsar. This model also assumes magnetic pair creation near the neutron star but posits that emission occurs under the influence of the surface $B$ (i.e., $\propto \sqrt{P \dot P}$). It is important to note that these analytical relations are simplifications of the complex processes occurring in pulsar magnetospheres \citep[][]{Harding2017}.  Therefore, definitive conclusions based on a precise comparison between these analytical relations and the observations are unwarranted.

We also sought to empirically determine the relationship between $E_{\rm SR}$ and the pulsar's spin parameters of the form $(E_{\rm SR}/E_{\rm SR,\ J1838})=C (P/P_{\rm J1838})^\alpha(\dot P/\dot{P}_{\rm J1838})^\beta$, where the subscript ``J1838'' means the values for J1838.  Our best-fit values for the exponents are $\alpha=-3.0\pm1.7$, and $\beta=1.7\pm0.7$. The measurements align with this trend almost perfectly, indicating overfitting. Furthermore, the substantial uncertainties associated with these exponents, stemming from a limited sample size, preclude meaningful constraints on the proposed scenarios. Further investigation with a larger dataset encompassing broader ranges of $E_{\rm SR}$, $P$, and $\dot P$ (i.e., including more MeV pulsars) is crucial to improve this empirical relationship.  Such studies will contribute to a deeper understanding of the physical processes governing synchrotron radiation in pulsar magnetospheres.
 
As discussed previously, the estimated $E_{\rm SR}$ values are subject to considerable uncertainty primarily due to their strong model dependence. The absence of observational data directly covering the $E_{\rm SR}$ energy range necessitates a large extrapolation of our models to higher energies. Consequently, minor inaccuracies in best-fit parameters, potentially induced by (cross-)calibration uncertainties, can propagate into large variations in the derived $E_{\rm SR}$ values. To resolve these challenges and enable more accurate $E_{\rm SR}$ estimations for MeV pulsars, deeper NuSTAR observations and future data from the COSI observatory \citep[][]{Tomsick2021} in the 0.2--5\,MeV band will be invaluable.

\subsection{X-ray Properties of G25.24 and AX~J1837}
\label{sec:sec4_2}
Our analysis of G25.24 revealed a softer $\Gamma$ of $2.1$ and a higher flux of $\fluxtt\approx 2.0\times 10^{-12}$\,\fluxcgs, compared with the previously reported values of $\Gamma=1.1\textrm{--}2.0$ and $\fluxtt=1.0\times 10^{-12}$\,\fluxcgs\ by \citet{Gotthelf2008}. We attributed these discrepancies primarily to differences in inferred $N_{\rm H}$ values, variations in extraction region sizes, and flux loss due to a chip gap (Section~\ref{sec:sec2_4_2}). We argue that our measurements provide a more accurate representation of the PWN emission due to the use of substantially larger datasets, including one unaffected by chip gaps. Moreover, our $N_{\rm H}$ estimates are consistent with high-quality pulsar data (Section~\ref{sec:sec2_4_1}) and independent studies \citep[e.g.,][]{Anada2009,Takata2024}.

We have provided an improved spectral characterization for AX~J1837 compared to previous studies. Our analysis found that the X-ray emission from this source is fainter and exhibits a harder spectrum than that of G25.24. However, our spectral measurements for both sources are associated with significant uncertainties due to the limited number of detected photons, a consequence of their large off-axis positions in the observations. This paucity of counts currently precludes precise spectral comparisons and accurate SED modeling. Therefore, deeper on-axis X-ray observations are essential for obtaining accurate spectral characterizations of these sources.

\subsection{VHE and UHE Emissions from the HESS~J1837$-$069 Region}
\label{sec:sec4_3}
As demonstrated in Section~\ref{sec:sec3_3}, a leptonic model incorporating both G25.24 and AX~J1837 is generally capable of reproducing the observed multi-wavelength emission from the HESS~J1837$-$069 region, including the broadband SED, the X-ray surface brightness profile of G25.24, and the spatial extent in the H.E.S.S. band, using the parameters detailed in Table~\ref{ta:ta4}.
However, this leptonic scenario does not represent the unique interpretation, given the physical complexity of the region (Figure~\ref{fig:fig1}). Other possibilities have been proposed in the literature.

\citet{Sun2020} attributed the Fermi-LAT and H.E.S.S. emissions observed from the HESS~J1837$-$069 region to G25.24 (associated with HESS~J1837$-$069) and the star cluster RSGC~1 (Figure~\ref{fig:fig1}), suggesting hadronic acceleration within the cluster. Alternatively, \citet{Katsuta2017} proposed G25.24 and the star cluster G25.18+0.26 as the origin of the gamma-ray emission. These prior studies utilized a H.E.S.S. SED measured over a smaller region \citep[][]{HESS2006} compared to our analysis \citep[using results of][]{HESSHGPS2018}. Consequently, the VHE SED models for HESS~J1837$-$069 presented in these previous works exhibit a flux level approximately three to four times lower than that adopted in our study; our PWN model (Table~\ref{ta:ta4} and Figure~\ref{fig:fig5}) could readily accommodate this lower VHE flux with reduced $U_{\rm IR}$.

While these star-cluster scenarios reproduce the Fermi-LAT and H.E.S.S. SEDs, they face several challenges. The Fermi-LAT emission regions (position and size) inferred by \citet{Katsuta2017} and \citet{Sun2020} show substantial discrepancies both with each other and with catalog values (4FGL and 3FHL), requiring further validation. RSGC~1 was suggested to be an inefficient hadron accelerator \citep[][]{Fujita2014}, and VHE emission from G25.18+0.26 appears subdominant (Figure~\ref{fig:fig1}). Furthermore, it remains unclear whether these models can adequately account for the X-ray emission from G25.24, the higher flux measured by a refined H.E.S.S. analysis \citep{HESSHGPS2018}, and the recent LHAASO UHE measurements (position and flux; Figures~\ref{fig:fig1} and \ref{fig:fig5}a).

On the other hand, our two-PWN scenario (Section~\ref{sec:sec3_3}), while plausible, currently lacks direct observational confirmation, specifically a firm identification of the pulsar powering AX~J1837. Furthermore, while our model accounts for the time evolution of energy injection, it does not fully incorporate the complex dynamical evolution of PWN properties (e.g., the model employs stationary $V_{\rm flow}$, $B$, and $p_1$). Therefore, confirmation through evolutionary PWN models is necessary to further validate this scenario.

Consequently, the current understanding of the VHE and UHE emissions from the HESS~J1837$-$069 region remains incomplete. It is conceivable that both the star clusters and the PWNe contribute to the observed gamma-ray emission. Future X-ray, VHE, and UHE observations are therefore essential to reach a definitive conclusion. 
\section{Summary}
\label{sec:sec5}
We conducted an X-ray analysis of the region around the MeV pulsar J1838, aiming to characterize the emission properties of J1838, G25.24, and the nearby PWN candidate AX~J1837. We then demonstrated the feasibility of a leptonic scenario for the VHE and UHE emissions observed from this region. Our key findings are summarized below:
\begin{itemize}
\item J1838 exhibits a curved X-ray SED with a peak at $73^{+85}_{-26}$\,keV estimated based on the {\tt logpar} model. Combining this measurement with data from other MeV pulsars, we constrain the scaling relation $E_{\rm SR}\propto P^{-3.0\pm1.7} \dot P^{1.7\pm0.7}$.  To refine both individual $E_{\rm SR}$ values and this scaling relation, obtaining data directly covering the SED peak energies through deeper NuSTAR and future COSI observations is essential.

\item We performed spectral analysis of the X-ray emission from G25.24 and AX~J1837. Although this analysis improved previous characterizations of the sources' spectra, our spectral measurements are subject to considerable uncertainties, and definitive conclusions require deeper on-axis X-ray observations.

\item We proposed that G25.24 and AX~J1837 are the primary sources of the observed VHE and UHE emissions from the HESS~J1837$-$069 region, respectively. This contrasts with models by \citet{Katsuta2017} and \citet{Sun2020} that implicate a star cluster (G25.18+0.26 or RSGC~1) as a substantial contributor to the region's gamma-ray emission. Future X-ray and higher angular resolution gamma-ray observations are crucial for definitively distinguishing between these competing scenarios.
\end{itemize}

\begin{acknowledgments}
We thank the anonymous referee for the constructive comments, which significantly improved the clarity of this paper. MP acknowledges support from Basic Science Research Program through the National Research Foundation of Korea (NRF) funded by the Ministry of Education (RS-2024-00462623).
JP acknowledges support from Basic Science Research Program through the National
Research Foundation of Korea (NRF) funded by the Ministry of Education (RS-2023-00274559).
This work was supported by the National Research Foundation of Korea (NRF) grant funded by the Korea government (MSIT) (RS-2023-NR076359). This work was supported by a funding for the academic research program of Chungbuk National University in 2025.

This paper employs a list of Chandra datasets, obtained by the Chandra X-ray Observatory, contained in Chandra Data Collection\dataset[DOI: 10.25574/cdc.357]{https://doi.org/10.25574/cdc.357}.

\end{acknowledgments}
\facilities{CXO, XMM, NICER, NuSTAR}
\software{
XSPEC \citep[v12.14.0;][]{a96},
HEAsoft \citep[v6.33;][]{heasarc2014},
CIAO \citep[v4.15;][]{CIAO2013},
XMM-SAS \citep[20230412\_1735;][]{xmmsas14},
}

\bibliography{ms}{}
\bibliographystyle{aasjournalv7}

\end{document}